\documentclass[twocolumn,showpacs,preprintnumbers,amsmath,amssymb]{revtex4}
\usepackage[english]{babel}

\usepackage{graphicx}

\newcommand{\be}[1]{ \begin{eqnarray} \mbox{$\label{#1}$} }

\newcommand{\ee}{\end{eqnarray}}
\newcommand{\pref}[1]{(\ref{#1})}

\newcounter{mycount}

\newcommand\ie {{\it i.e. }}
\newcommand\eg {{\it e.g. }}

\newcommand\etcp{{\it etc.. }}

\newcommand\ket [1] {|#1 \rangle }

\newcommand{\av}[1]{\langle #1\rangle}

\newcommand\noi{\noindent}

\begin{document}

\title { Topological Field Theory for $p$-wave Superconductors  }

\author{T.H. Hansson$^1$}
\author{A. Karlhede$^1$}
\author{Masatoshi Sato$^2$}
\affiliation{$^1$Department of Physics, Stockholm University, AlbaNova University Center, SE-106 91 Stockholm,
Sweden \\
$^2$The Institute for Solid State Physics, 
The University of Tokyo, Kashiwanoha 5-1-5, 
Kashiwa-shi, Chiba 277-8581, Japan }

\date{\today}

\begin{abstract} 

We propose a topological field theory for a spin-less two-dimensional  chiral superconductor that contains fundamental Majorana fields. Due to a fermionic
gauge symmetry, the Majorana modes survive as dynamical degrees of freedom only at magnetic vortex cores, and on edges. 
We argue that these modes have the topological properties pertinent to a $p$-wave superconductor including the non-abelian braiding statistics, and
support this claim by calculating the ground state degeneracy on a torus. 
We also briefly discuss the connection to the Moore-Read Pfaffian quantum Hall state, and extensions to the spinful case and to three-dimensonal topological superconductors. 

\end{abstract}

\pacs{74.20.De, 74.20.Rp, 03.65.Vf}

\maketitle

{\em Introduction} -- The great current interest in topological phases of matter is mainly due to the intrinsic scientific challenges to 
discover, engineer and describe them.
The hallmark of a topological phase is that the elementary excitations
carry unusual quantum numbers, and that the quantum mechanical wave
functions acquire phase factors that are insensitive to noise, when these
excitations are braided around each other. It has been proposed to code
quantum information in these phase factors, with the prospect  to
greatly improve both quantum memories and quantum processors\cite{comp};
this possible technological application is fascinating. 
The field was initiated by the discovery of the quantum Hall (QH) effect, 
fueled by the theoretical proposal for topological spin liquids,
and recently by the surge of investigations of topological insulators and
topological superconductors (SC)\cite{reviews}. The theoretical study of these systems range from
detailed microscopic models based on realistic band structure, to
general descriptions in terms of effective low energy theories. An
important class of the latter are the topological field theories (TFTs)
that distill the topological information about the states of
matter. Prime example of such TFTs are the Chern-Simons (CS) theories
that describe the incompressible quantum Hall fluids.  

The perhaps most familiar system that is a topological phase is an ordinary
$s$-wave SC, where the non-trivial braiding phase is the minus
sign picked up by the wave function as a spinon (a neutral spin half
Bogoliubov quasiparticle) encircles a vortex of strength $h/2e$ in the
Cooper pair condensate\cite{kivrok}.  That this is a {\em bona fide}
topological phase was stressed by Wen\cite{wensup}, and it was later
realized that  the corresponding topological field theory is a $BF$
theory, which in two space dimensions is closely related to the CS
theories\cite{lein,hos}. For the case of three space dimensions it 
was known that the BF theories could provide a topological mass\cite{allen},
and the connection to superconductivity was discussed in several 
papers\cite{bal}.
A $BF$ theory  for $d$-wave SC was
conjectured\cite{hos}, and derived and analysed by
Hermanns\cite{mariathesis}. Recently it has  also been shown that
the TFTs describing topological insulators are of the $BF$
type\cite{joel}. It is thus natural to assume that also a chiral $p$-wave
superconductor should be described by a $BF$ theory, but it is also
obvious that such a theory must be more complicated in order to
correctly describe the gapless Majorana modes known to exist both in
vortex cores, and at boundaries\cite{readgreen,fendley}.  Furthermore, the
vortices must satisfy the braiding rules appropriate for Ising type
non-abelian statistics\cite{mr,NW,ivanov,fendley}. In this letter we
propose a TFT for a two-dimensional  spin-less chiral SC
that satisfies these requirements, comment upon the relation to the Moore-Read Pfaffian
QH state, and briefly touch upon  how to
incorporate spin and how to generalize to the three-dimensional case.  

Our theory differ from previously proposed  effective theories  for 
the Moore-Read state in that it has a fundamental Majorana fermion field 
that enters at the same level as the topological gauge fields, and just
as these it does not have any bulk dynamics. In the same way  as a Chern-Simons
gauge field, it does have dynamics on edges, but also at vortex 
cores, which for many purposes can be modeled as internal boundaries. 
The rationale for this picture is that it explicitly exhibits the full low 
energy content of the theory. This is in contrast to  the approach based on 
non-abelian Chern-Simons theory\cite{fradnayak},  where the presence 
of Majorana zero modes enters implicitly by the choice of representation 
of the Wilson loops that describe the worldlines of the fundamental 
quasiparticles. 

{\em The model} -- Our model is defined by 
\be{part} 
Z[j_{\rm v},j_{\rm q}] 
= \int { \cal D} [a_\mu] {\cal D} [b_\mu]  {\cal D} [\gamma] 
e^{i\int d^3 x\, {\cal L} }
\ee 
and
\be{lag}
{\cal L}
&=&\frac{1}{\pi}\epsilon^{\mu\nu\rho} [ b_{\mu}\partial_{\nu}a_{\rho} 
+  \frac i 4 a_{\mu}\partial_{\nu}\gamma \,
\partial_{\rho}\gamma  ]    -  j_{\rm q}^{\mu}a_{\mu}-j_{\rm v}^{\mu}b_{\mu}  +  {\cal L}_{\rm{gf}}
\nonumber \\
&=& \frac{1}{\pi}\epsilon^{\mu\nu\rho}
\left( b_{\mu}+ \frac 1 4 \gamma \, i \partial_{\mu}\gamma \right)\partial_\nu a_\rho 
-  j_{\rm q}^{\mu}a_{\mu}\nonumber \\
&-&   j_{\rm v}^{\mu}b_{\mu}   -    \frac{i}{4\pi}\epsilon^{\mu\nu\rho}\partial_{\mu}\left[
a_{\nu}\gamma\, \partial_{\rho}\gamma\right]  + {\cal L}_{\mathrm {gf}} \, ,
\ee
where the second line is obtained by partial integration.  
$j_{\rm v}$ and $j_{\rm q}$ are vortex and quasiparticle currents
respectively, 
$a$ and $b$ are gauge potentials,  and $\gamma$ is a (one-component) Majorana fermion. 
${\cal L}_{\mathrm {gf}}$ is  a gauge fixing term that is discussed below. 
Including only the first term in the parenthesis, we retain the $BF$
action, $\sim bda$, for an $s$-wave SC.
Under parity, $P$,  $(x,y)\rightarrow (-x,y)$, the fields in \pref{lag}
transform as  $ (a_0,a_x,a_y) \rightarrow (a_0,-a_x,a_y)$, 
$ (b_0,b_x,b_y) \rightarrow (-b_0,b_x,-b_y)$, and $\gamma \rightarrow \gamma$ and 
under time reversal, $T$,  as
$ (a_0,a_x,a_y) \rightarrow (a_0,-a_x,-a_y)$, 
$ (b_0,b_x,b_y) \rightarrow (-b_0,b_x,b_y)$ and  $\gamma \rightarrow
\gamma$. As is well known, this implies that $bda$ is invariant under both
$P$ and $T$, while the  new term $i\gamma d\gamma da$ violates 
both these symmetries as appropriate for  a chiral SC.  

We now give  a heuristic argument, which will be made more precise later, for why \pref{part} and \pref{lag}
describe a SC with dynamical Majorana modes localized on vortices and edges.
Up to boundary terms, \pref{lag} is invariant under the two gauge
transformations, $a_\mu \rightarrow a_\mu + \partial_\mu \Lambda_a$ and
$b_\mu \rightarrow b_\mu + \partial_\mu \Lambda_b$, and it is known that in the $s$-wave case the boundary supports bosonic  degrees of freedom that
will generically be gapped\cite{hos}. In the present case, this is not true as will be discussed later. 
The terms in the second line of  \pref{lag} possess another local symmetry, 
\be{susy}
\gamma &\rightarrow & \gamma + 2 \theta,  \\
b_\mu &\rightarrow &b_{\mu}-i\theta \partial_{\mu}\gamma/2-i\gamma\partial_{\mu}\theta/2-i\theta\partial_{\mu}\theta  \ ,
\nonumber 
\ee 
where $\theta$ is a Majorana fermion field. 
This symmetry is violated by the terms on the third line of 
\pref{lag}, \ie by the total derivative term and the one proportional to the vortex current.
This implies that the Majorana fermion can be gauged away everywhere  except 
at the boundary and at vortex cores, where it emerges as a dynamical degree of freedom,
in close analogy to the bosonic edge modes in CS or BF theories\cite{wen,hos}.
Using the suggestive notation
\be{btilde}
\tilde b_\mu = i \gamma\partial_\mu \gamma/4 \, ,
\ee
we note that \pref{lag} is quadratic in $a$,
$b$ and $\tilde b$, and linear in derivatives.

Integrating the Lagrange multiplier field $b$, yields $da = \pi j_{\rm v}$,
showing that a unit strength  vortex carries a flux $\pi$. 
Using the notation $\frac \pi d $ for the pertinent Green's function, we have
 $a = \frac \pi d j_{\rm v} + d \Lambda_a  $
and substituting this in \pref{lag} gives the bulk Lagrangian
\be{lag2}
{\cal L} = \frac 1 4 j_{\rm v}^\mu  \gamma i\partial_\mu\gamma   -
j_{\rm q}^{\mu}a_{\mu} 
= {\cal L}_{\rm maj} - 
j_{\rm v} \frac
\pi d j_{\rm q} \, ,
\ee
where we ignored the gauge-fixing term.
The last term in \pref{lag2} is an abelian  statistical interaction that gives a  $\pi$ phase for a
vortex circling a quasiparticle. If there were no
internal structure to the vortices, \ie no Majorana modes, two vortices
with the same charge would be identical bosons. 

To exhibit the  Majorana modes, we consider a vortex source corresponding to $N$  Wilson
loops $W_{C_a}=\exp(i\oint_{C_a} dx^\mu b_\mu)$,
\be{vsuorce}
j_{\rm v}^\mu(x^\mu)=\sum_{a=1}^{N}  \int_0^1 d\tau\,  \delta^3(x^\mu-x_a^\mu(\tau)) \frac { x^\mu_a (\tau)} {d\tau}\ ,
\ee
where the loop $C_a$ is parametrized as $x^\mu_a(\tau)$. Substituting in \pref{lag2} yields the action
\be{majact} 
{\cal S}_{\rm maj} =
\frac 1 4 \sum_{a=1}^{N} \int_0^1 d\tau\, 
{\gamma}_a(\tau)
i\partial_{\tau}{\gamma}_a(\tau) \, ,
\ee
where $\gamma_a(\tau) \equiv \gamma(x^\mu_a(\tau))$. 
For the simple case of the loops extending  over all times, and crossing any fixed time surface only twice,  
we can use the time, $t$, as a parameter to rewrite this as,
\be{edact} 
{\cal S}_{\rm maj} =
\frac 1 4 \sum_{a=1}^{2N} \int_{-\infty} ^\infty dt\, 
{\gamma}_a(t)
i\partial_t{\gamma}_a(t) \, ,
\ee
which, as promised, describes a collection of gapless Majorana
fermions living on the world lines of the vortices.

{\em Braiding and statistics} -- To demonstrate that \pref{edact}
 reproduces the Ising braiding statistics of the vortices, we
switch to an operator description, where  \pref{edact} implies the 
canonical commutation relations,
\be{cancom}
\{ \hat{\gamma}_a(t), \hat{\gamma}_b(t) \} =   2\delta_{ab} \, .
\ee
Since there is an even number of vortices  we can form the
complex Dirac fields
$
\hat \psi_A 
=  (\hat{\gamma}_{2A-1}+i\hat{\gamma}_{2A}) / 2 
$,
where  $A = 1,2 \dots N $,
which together with $\hat \psi_A^\dagger$ satisfy 
$\{\hat \psi_A^{\dagger},\hat \psi_B\} = \delta_{AB}$ and
$\{\hat \psi_A,\hat \psi_B\} = \{\hat \psi^{\dagger}_A,\hat \psi^{\dagger}_B\}=0.$
The pairing of Majorana fermions in the Dirac fields is arbitrary, and amounts to a choice of basis in the Hilbert space spanned by the vectors
$
|  \alpha_1 \cdots \alpha_{N} \rangle
=   |\alpha_1\rangle\otimes\cdots\otimes
|\alpha_{N}\rangle,  
$
where $\alpha_A=0,1$ and $\hat \psi_A^{\dagger}|0\rangle=|1\rangle$ and
$\hat \psi_A|1\rangle=|0\rangle$. The full description of the low-energy sector, in the absence of quasiparticles, 
is thus in terms of the state vectors, $\ket { \vec x_1 \dots \vec x_{2N} ; \alpha_1 \cdots \alpha_{N} }$, 
since the sources determine the gauge potentials $a(\vec x,t)$ and  $b(\vec x,t)$ up to gauge
transformations. 
Also note that the Hamiltonian corresponding to the action \pref{edact} is identically zero, so the 
time evolution is entirely determined by Berry matrices.
From general principles, we also know that these statistical matrices will only depend on the
braiding patterns, since they will constitute a (non-abelian) representation of the braid group.
This  group is  generated by 
the elementary exchanges ${\mathrm T_a}: \vec x_a  \leftrightarrow \vec x_{a+1}$, that
do not involve any braiding with the other vortices, and
these exchange processes can be viewed as time evolutions, 
\be{timeev}
U(T) \ket { \vec x_1 \dots \vec x_a, \vec x_{a+1} \dots \vec x_{2N} ; \alpha_1\cdots \alpha_{N} }_0  \\
= \sum_{\{\alpha' \}} C_{\alpha'_1\dots \alpha'_{N/2}}
 \ket { \vec x_1 \dots \vec x_{a+1}, \vec x_{a} \dots  \vec x_{2N} ; \alpha'_1 \cdots \alpha'_{N} }_T \, .  \nonumber
 \ee
where the unitary operator $U(T)$   interchanges the position of the $a$-th and $(a+1)$-st vortex, 
leaving the others unchanged. We assume that the Majorana operators, $\hat\gamma_a$, 
which are attached to the vortices, are interchanged when the corresponding vortices are interchanged, 
$U(T)\hat \gamma_a U^\dagger (T) = \zeta_a  \hat \gamma_{a+1}$ and
$U(T)\hat \gamma_{a+1}U^\dagger (T) = \zeta_{a+1} \hat \gamma_{a}$. 
Since $\hat \gamma_a^\dagger=\hat \gamma_a$, and the commutations relations should be preserved,  
we must have  $\zeta_a = \pm 1$, but any combination of these signs is allowed.  
There are, however, only two unitary bosonic\footnote{
Fermionic parity is preserved under time evolution so $U$ must be unitary and bosonic.}
operators  that  exchange $\hat \gamma_a$ and $\hat \gamma_{a+1}$, 
while leaving all other $\hat \gamma$s unchanged. These are 
$U_\pm = e^{\pm \frac \pi 4 \hat \gamma_a\hat \gamma_{a+1}}$, corresponding to $\zeta_a= 
-\zeta_{a+1} = \mp 1$, which are precisely the transformations that imply Ising non-abelian 
statistics as shown by Wilczek and Nayak in the context of the Moore-Read QH state\cite{NW}, and
by  Ivanov in the context of a Bogoliubov-de Gennes description of a $p$-wave SC\cite{ivanov}. 
If $\hat \psi_a = (\hat \gamma_a +i \hat \gamma_{a+1})/2$, then
$U_\pm  = e^{\pm i \frac \pi 4} e^{\mp i \frac \pi 2 \hat \psi_a^\dagger \hat \psi_a}$,
implying \eg  $U_+\ket 0 = e^{i\frac \pi 4}\ket 0$ and $U_+\ket 1 = e^{-i\frac \pi 4}\ket 1$, where
$\hat \psi_a^\dagger \hat \psi_a\ket n = n\ket n$. That $U$ is diagonal in this basis follows already from conservation 
of fermionic parity, so the non-trivial result giving the non-abelian statistics is the values of the phases. 
In Ivanov's treatment, the crucial relative sign, $\zeta_a = -\zeta_{a+1}$
originates from the coupling of  the Bogoliubov-de Gennes quasiparticle  
to the superconducting order parameter, and we believe that 
a similar interpretation is valid also here, since the term $\sim a d\tilde b$
in the topological action identifies $\epsilon^{ij}\partial_i\tilde b_j$ as 
a $Z_2$ quasiparticle charge. 
The above argument for non-abelian statistics
is based solely on the general properties of $U_\pm$, and it would certainly be
interesting to derive the expression for $U_\pm$ directly by considering time evolution
in the topological field theory. By  quantizing the full 
field theory using canonical methods, one could determine the 
time evolution of the Majorana fields 
$\gamma_a$, and from there the time evolution of the state vectors. This is, however,
quite complicated due to the singular nature of the constraints.

  {\em Ground state degeneracy on a torus} --
To further strengthen the case for non-abelian statistics we now show that, using a natural lattice regularization, there are 
three degenerate bosonic ground states on the torus.

 Following  the strategy of  Refs. \cite{hos} and
\cite{oksnt}, we solve the constraints in the absence of sources as, $a_i=\partial_i \Lambda_a+\overline a_i/L_i$,
$\beta_i = b_i+\tilde b_i=\partial_i \Lambda_\beta+\overline \beta_i/L_i$, where $\Lambda_{a/\beta}$ are periodic 
and $\overline a_i, \, \overline\beta_i$ are constant in space; $L_i$ is the length of the torus in direction $i$, 
so $\overline a_i = \int_{C_i} dx^i a_i$ and $\overline\beta_i =  \int_{C_i} dx^i \beta_i $, where the  contour  $C_i$ winds the 
torus once in the $i$-th direction, but is otherwise arbitrary. Note that $\beta_i$ is invariant under the fermionic gauge 
transformation.
We get the Lagrangian $L$ as $\pi L=\epsilon^{ij}\dot {\overline a}_i \overline\beta_j$, which implies the commutation relations, 
$[\overline a_i, \overline \beta_j] = i \pi \epsilon^{ij}$. Next we define the operators,
\be{gsops}
{\cal A}_i &=& e^{i\, \overline a_i}   \\ 
{\cal B}_i &=& \int_ {\gamma \in C_i}{\cal D}\gamma\,  e^{i \overline \beta_i } =
  \int_ {\gamma \in C_i}{\cal D}\gamma\,  e^{i[ \overline b_{i}  +  \frac i 4  \int_{C_i} dx^i   \gamma  \partial_i \gamma] }  \nonumber\, , 
\ee
with $\overline b_i=\int_{C_i}dx^i b_i$,
which   satisfy 
$\{ {\cal A}_x,  {\cal B}_y \} = \{ {\cal A}_y,  {\cal B}_x \} =0$, since $\overline a_i$ and $\overline\beta_j$ obey the same commutation
relations as $\overline a_i$ and $\overline b_j$ in the $s$-wave case. 
We would thus expect that, just as in that case, there would be a 4-fold degeneracy of the ground state as a direct consequence of the commutator algebra.
The situation here is however more subtle, since $\overline \beta_i$ depends on the fermionic variable and
the fermionic integration in ${\cal B}_i $ is necessary to obtain an operator that acts only in the bosonic part 
of the Hilbert space only. A direct calculation shows that for ${\cal B}_i$ to
be non-zero, we must use anti-periodic boundary conditions on $\gamma$ in the $i$-th direction, which is 
consistent with $\gamma$ being fermionic.
That we need the fermionic integration in \pref{gsops} can also be understood from the Lagrangian point of view. If we try to calculate a correlation function including the naive operator $e^{i\overline b_i}$, we notice that this is not gauge invariant, and we cannot 
gaugefix the Majorana field on the contour $C_i$. We can thus naturally associate this integral with the operator, and to have a gauge invariant object we must also add the $\tilde b_i$ contribution to get ${\cal B}_i$. 

The new complication occurs when we attempt to explicitly construct the four ground states.
 Since ${\cal B}_i$ measures the $a_j$ flux, one
 would  naively expect that the four degenerate ground states, just as in the $s$-wave case, should be labeled
 by thse fluxes. Denoting the
 ${\cal A}_i$ eigenstates by $\ket {s_x,s_y}$ we would have $\ket {+,+}$, $\ket{+,-} = {\cal B}_y \ket{+,+}$, 
 $\ket{-,+}={\cal B}_x \ket{+,+}$ and $\ket{-,-}={\cal B}_x{\cal
 B}_y\ket{+,+}$. In our case, this is no longer true. The  state $\ket {-,-} $
 is absent since the operator $  {\cal B}_x {\cal B}_y $ vanishes due to the fermionic integration, and we
 are left with three ground states as expected if the excitations obey non-abelian statistics. Heuristically,
 this "blocking mechanism"\cite{oksnt} comes about since the two contours $C_x$ and $C_y$ necessarily cross, which amounts to having two fermions at the same point. A more rigorous argument is given below. 
 
{\em Discrete formulation and fermionic gauge fixing} --
 To prove the above statements about the operators ${\cal B}_i$, 
 and also to explain  the fermionic gauge-fixing, we extend  the discretized 
BF theory due to Adams\cite{adamsBF} to our Lagrangian \pref{lag}. In Adams' construction, 
the gauge potentials are real-valued one-forms, $a$ and $b$,
defined on the links of two dual cubic lattices $K$ and $\hat K$. The action is 
\be{disact}
S = \frac 1 \pi \int d^3x\,  b \star^K d^K a 
\ee
where $\star^K$ is a duality operator that relates a $p$-form on $K$ to a $(3-p)$-form on $\hat K$, \ie a link on $K$ to 
a plaquette on $\hat K$ \etcp 
The coboundary operator, $d^K$ , is the discretized version of the outer
derivative $d$, so one can thus think of \pref{disact} as a sum of products of a link on the $K$ lattice with
the dual plaquette it pierces.  
Adams showed that \pref{disact} is invariant under $a\rightarrow a + d^K \Lambda_a$ and 
$b\rightarrow b + d^{\hat{K}} \Lambda_b$, and that it  reproduces the expectation values of Wilson loops 
as calculated in the continuum theory. Using standard arguments,  this also implies that 
the commutation relations among the operators ${\cal A}_i$ and ${\cal B}_j$ are the same as in the continuum.
Also, the coboundary operator, $d^K$, satisfies the Liebniz rule, so partial integrations
can be performed as usual\cite{adamspc}.

The vortex  Wilson loop $W_{C}$ has the discrete
 counterpart $\exp(i\sum_{\langle ij \rangle\in C} b_{ij})$, where
$C$ is a loop on the dual lattice $\hat K$, and $b_{ij}$ the gauge one-form on the link $\av{ij}$.
We can  extend this construction to give a discretized version of our action by simply
making the replacement $b_{ij} \rightarrow \beta_{ij} =  b_{ij} + \frac i 4\gamma_i  \gamma_j$ where $\gamma_i$ is a Majorana fermionic 0-form defined on
the sites of the dual lattice $\hat K$. The discrete version of the fermionic gauge symmetry easily follows.

It is now clear that we have to gauge fix
all the $\gamma_i$, except those living on the loops, and  the simplest
procedure is to introduce the factor 
$
{\cal G}(\gamma_i)  = \prod_{i \notin \cup_{C}}     \delta(\gamma_i) 
$,
where $\delta(\gamma_i) = \gamma_i$ is a fermionic delta function, 
in the path integral measure.

 With this we have given a precise prescription for how to introduce point-like vortices,
and shown that they will support dynamical Majorana degrees of freedom.
 
The discrete form of the operators ${\cal B}_i$ directly follows,
\be{Bdisc}
 {\cal B}'_i = \int  \prod_{\gamma_k \in C_i} d\gamma_k  \,   e^{i  \sum_{\av{ij}\in C_i}(  b_{ij} + \frac i 4\gamma_i \gamma_j) } \, ,
\ee
where we note that the field in the exponent is the gauge invariant combination $\beta_{ij}$ defined above.
To evaluate the fermionic integrals, we first assume that there is an even number of sites in both the $x$ and $y$ directions.
It is then clear that both ${\cal B}_x'$ and ${\cal B}_y'$ are non-zero, since all sites on the loop can be covered  once, and
only once, by expanding the exponential. The proper discretized form of the product operator is,
\be{BBdisc}
( {\cal B}_x  {\cal B}_y)'= \int  \prod_{\gamma_k \in C_x \cup C_y } d\gamma_k    
e^{i  \sum_{\av{ij}\in   C_x \cup C_y }(  b_{ij} + \frac i 4\gamma_i \gamma_j) } \, .
\ee
This operator vanishes since there is an odd  number of fermionic 
 integrals and expanding the exponential will
only bring down an even number of $\gamma$s. Similarly, for an even - odd lattice, ${\cal B}_x'$ and $( {\cal B}_x  {\cal B}_y)'$ will
give new states, while ${{\cal B}^\prime_y}$ vanishes, \etcp  It is also easy to see that if we allow for fermionic  states, corresponding to an 
odd number of vortices, there will be an extra ground state\cite{readgreen}.

{\em Edge states} -- 
We already stressed that the total derivative term in \pref{lag} violates
the fermionic gauge symmetry and we can directly read off the action
for the  dynamical Majorana edge mode,
\be{edgeact}
S_{\rm ed} = \frac i {4\pi}  \int dx dt \,  \gamma (x,t)  [ a_x  \partial_0 - a_0
\partial_x] \gamma(x,t)  \, ,
\ee
where $x$ is the coordinate along the edge. 
The line integral $\oint dx\, a_x$ measures the total vorticity in the system
which can have both a bulk and an edge contribution. In a $p$-wave
superconductor an edge vorticity, corresponding to an edge current, 
is indeed expected. In our topological description this amounts to 
taking $a_x$ as an unknown constant. The value of $a_0$ is formally a gauge
choice, but this is somewhat misleading since it in fact determines 
the velocity $v = a_0/a_x$ of the edge modes and thus is related to the edge potential.
The situation is very similar to that for the edge modes of QH liquids, as
discussed by Wen\cite{wen}. For a circular droplet with radius $R$, 
the energy of the fermionic edge modes
is $E = vl/R$ where $l$ is the angular momentum, which is integer or half-integer
valued depending on whether the boundary conditions are periodic  or 
anti-periodic. For consistency, we must thus take periodic boundary conditions if there is 
an odd number of vortices in the bulk, and antiperiodic if the number is even. 
In this way the zero angular momentum edge mode will provide the missing
Majorana zero mode, which combined with the localized bulk Majorana modes in a vortex form
Dirac fermions. In a careful treatment of the gauge fixing, these boundary conditions
are likely to be necessary for the fermionic path integral not to vanish. 

{\em External fields and the Moore-Read QH state} --
Coupling to an external electromagnetic potential, $A$, is obtained by shifting
$a \rightarrow a+ A$ as explained in  Ref. \onlinecite{hos}. 
From the work of Reed and Green, we expect that our theory should also
describe the  non-abelian Moore-Read, or Pfaffian, QH state which 
can be viewed as a $p$-wave paired state of composite fermions\cite{GW,readgreen}.

A crucial difference between this state and the $p$-wave SC is that  flux and charge cannot be separated, 
so the charge $e/4$ qusiparticle current couples to both $a$ and $b$, 
\be{mr} 
{\cal L}_{{\rm MR}} &=&
\frac{1}{\pi}\epsilon^{\mu\nu\rho} ( b_{\mu} + \tilde b_\mu )  \partial_{\nu} (a_{\rho} +  \frac e 2 A_\rho) \\
   &+&  \frac e {4\pi} \epsilon^{\mu\nu\rho} a_\mu \partial_\nu  A_\rho  -  j^{\mu}  
 (\frac 1 4 a_{\mu} +\frac 1 2 b_{\mu} )        \, , \nonumber
\ee 
where we also added the coupling to $A$ and   omitted the total derivative and the gauge-fixing terms.
To get a physical motivations for the coupling terms in \pref{mr}, we note that the bosonic part
can be written on the standard form\cite{wen},
\be{coupl}
 \frac 1 {2\pi}  \epsilon^{\mu\nu\rho} t_I  c_\mu^I \partial_\nu A_\rho - l_I j ^\mu c^I_\mu \nonumber
 \ee
where $c^1 = a$,  $c^2 = b$, and  $\mathbf t = (1/2, 1)$, $\mathbf l= (1/4, 1/2)$.
We see that the $l_I$ vector is chosen to describe a Moore-Read quasiparticle which is 
a half vortex with charge $e/4$. The second component in the charge vector $t_I$ just 
expresses that $A_\mu$ couples to the charge current of the condensate, $\sim db$, 
with unit strength, while the factor 1/2 in the  first component  expresses the flux 
charge connection in the quantum Hall fluid. The unit of vorticity in the vortex condensate,
$\sim db$, is a unit (or Laughlin) vortex whcih carries a charge given by the filling fraction $\nu=1/2$.  

Following the same steps as took us from \pref{part} to \pref{lag2}, 
\be{MR2}
{\cal L}_{{\rm MR}} = 
 {\cal L}_{\rm maj}   - \frac {e^2} {8\pi} AdA + \frac e 4 jA - \frac 18 j \frac \pi d j         \, ,
\ee
where ${\cal L}_{\rm maj} $ differs from that defined in \pref{lag2} by a factor of 2 and the presence of
a coupling $\sim \tilde b dA$. 
The second term in \pref{MR2} gives the correct Hall conductivity,  the third term shows that the quasiparticles have charge
$e/4$, and the  last term  provides them with the  abelian statistical angle $\pi /8$, which in the 
 conformal field theory description originates from the charge sector\cite{mr}. 
Thus the Lagrangian \pref{mr} encodes all the pertinent properties of the MR state. 

We can add local source term with the $\mathbf l$ vector $\mathbf l_{\mathrm {Lh}} = (1/2, 1)$ to describe  the charge $e/2$,
semionic, Laughlin holes, but it is less clear how to incorporate the neutral fermions except as composites of
the non-abelian quasiparticles. 
Finally, we note that the coupling of the current $j$ to both $a$ and $b$ has a natural 
counterpart in the discretized theory\cite{adamsBF}.

{\em Generalizations} -- The generalization of  \pref{part} and
\pref{lag} to include spin, and to $T$-invariant topological SCs such as the
proposed 2D $p$-wave Rashba non-centrosymmetric SC\cite{sato}, can be
obtained by combining several copies of \pref{lag}, as will be described
in future work\cite{hmk 2}.
A 3D $s$-wave SC is described by a TFT of point-like quasiparticles and
string-like vortices \cite{bal}, and a natural
generalization for a 3D $T$-invariant topological SC is,  
$
S_{\rm top} = \int d^{4}x\,  \left[  {\cal L}_{BF} -    j_{\rm qp}^{\mu}a_{\mu}-
J_{\rm v}^{\mu\nu}b_{\mu\nu}      \right] ,
$
where 
\be{bf3}
{\cal L}_{BF} =
\frac 1 \pi \epsilon^{\mu\nu\sigma\lambda}
(b_{\mu\nu}  + \tilde b_{\mu\nu} )   \partial_\sigma a_\lambda \, .
\ee
Here the antisymmetric tensor gauge potential  $b$ couples to the world
sheets of the strings defined by the source $J_{\rm v}^{\mu\nu}$\cite{bal,hos}.
The new term  involves the potential $\tilde b_{\mu\nu} $, which is defined in
analogy with \pref{btilde} but  this time 
in terms of {\em two} Majorana fields $\gamma_1$ and $\gamma_2$ as,
\be{3d bitilde}
\tilde b_{\mu\nu} = \frac{i}{2}\gamma^T \sigma_\mu \partial_\nu \gamma \ ,
\ee
where $\gamma^T = (\gamma_1, \gamma_2)$, and $\sigma_\mu = ({\mathbf 1},
\sigma_i)$, with $\sigma_i$ the Pauli matrices.  Analogously to the 2D
case, there is a local fermionic symmetry which implies that 
the Majorana fermions live only on the world sheets of the
strings, and for the simplest case of a stationary, circular string with
radius $R$ parametrized by the angle $\varphi$, we get the action,
\be{string}
S_{\rm string} = \frac 1 {2\pi} \int dt  \int_0^{2\pi} d\varphi \, \gamma^T
\left(i\sigma_\varphi \partial_t - {\mathbf 1} \frac {i\partial_\varphi}
R \right) \gamma
\ee
with $\sigma_\varphi = \cos\varphi\sigma_y - \sin\varphi\sigma_x $. This
action describes two 1D Majorana fields living on the vortex.

{\em Summary and outlook} -- We have proposed a topological action,
involving Majorana fields, 
 for a two-dimensional  spin-less chiral SC that contains fundamental Majorana fields. This
action possesses a local fermionic symmetry such that the Majorana
fields are pure gauge degrees of freedom except at vortex cores and on
boundaries. We have identified these dynamical Majorana degrees of freedom and shown
that they reproduce the known topological features of chiral SCs,
including the non-abelian braiding statistics, and the three-fold ground state 
degeneracy on the torus. 
We also gave a discrete version of our model, where the concept of
vortex core is sharply defined, and where the necessary gauge fixing
procedure could be unambiguously defined, thus justifying 
the more formal treatment in the continuum model.

There are several interesting directions to pursue. 
We already mentioned the problem of canonically quantizing the theory, 
where one should be able to explicitly calculate the 
non-abelian braiding phases,  the extension to the
spin-ful case where the details should be worked out, and also the
problem of how to consistently include an edge vorticity to describe the
expected edge  currents. 

In addition one should clarify the relation between our effective theory by \pref{mr}
for the Moore-Read state QH state, and the low-energy theory based on a 
$SU(2)_2$ Chern-Simons thory previoulsy proposed in Ref. \cite{fradnayak}.
Also, using the discrete model, it should be possible to  calculate the expectation values
of linked vortex loops, from which one can extract the non-abelian statistics 
as is done in \cite{mr}.  
A natural way to make the continuum theory \pref{lag} less singular is
to add a (non-topological) Maxwell terms for the gauge fields. As explained in 
\cite{hos}, this gives a finite London length and a dynamic plasmon mode. 
We expect that this will introduce gapped fermion modes in the vortices, but that a single
gapless Majorana mode  remains; this remains to be shown. \\

Although the fermionic gauge symmetry \pref{susy} 
is not a supersymmetry, it is nevertheless interesting to
speculate that there might be a connection to supersymmetric models. An obvious
candidate to study would be the supersymmetric extension of the relativistic
abelian Higgs model that gives a microscopic derivation of the $BF$ theory for $s$-wave SCs \cite{bal,hos}.

Another  challenging task is to investigate the 3D models of the
type \pref{bf3}, especially concerning edge modes.
Also, since the Majorana modes live on strings,  there might be a
connection between the lattice version of \pref{bf3} and
the Kitaev chain\cite{kitaev}, which is known to support Majorana modes
at the ends of open strings. 

\noi
{\bf Acknowledgement:} We thank  Fawad Hassan, Paramesweran Nair, Alexis 
Polychronakos, Shivaji Sondhi for helpful  discussions, and D.H.  Adams for 
explaining some results in Refs. \cite{adamsBF} and \cite{adamspc} to us.
 This work was supported by the Swedish
Research Council, and the Grant-in Aid for Scientific Research from MEXT
of Japan ``Topological Quantum Phenomena'' No.22103005 and No.22540383.

\end{document}